\documentclass[12pt]{article}
\usepackage[paper=a4paper,margin=1in]{geometry}
\usepackage[latin1]{inputenc}
\usepackage[T1]{fontenc} 
\usepackage{amsmath}
\usepackage{amsfonts}
\usepackage{amssymb}
\usepackage{amsthm}
\usepackage{graphicx}
\usepackage{mathrsfs}  

\numberwithin{equation}{section}

\begin{document}
\bibliographystyle{unsrt}

\title{A note on the post-Newtonian limit of quasi-local energy 
expressions}

\author{J\"org Frauendiener \\
Department of Mathematics and Statistics, University of Otago, \\
P. O. Box 56, Dunedin 9010, New Zealand\\
and Centre of Mathematics for Applications, University of Oslo,\\
P.O. Box 1053, Blindern, NO-0316 Oslo, Norway
\and
L\'aszl\'o B. Szabados \\
Research Institute for Particle and Nuclear Physics, \\
H--1525 Budapest 114, P. O. Box 49, Hungary}
\maketitle

\maketitle

\begin{abstract}
An `effective' quasi-local energy expression, motivated by the 
(relativistically corrected) Newtonian theory, is introduced in 
exact GR as the volume integral of all the source terms in the 
field equation for the Newtonian potential in static spacetimes. 
In particular, we exhibit a new post-Newtonian correction in 
the source term in the field equation for the Newtonian 
gravitational potential. In asymptotically flat spacetimes this 
expression tends to the ADM energy at spatial infinity as a 
{\em monotonically decreasing} set function. We prove its 
positivity in spherically symmetric spacetimes under certain 
energy conditions, and that its vanishing characterizes flatness. 
We argue that any physically acceptable quasi-local energy 
expression should behave qualitatively like this `effective' 
energy expression in this limit. 
\end{abstract}


\section{Introduction}
\label{sec-1}

In non-gravitational classical field theories on flat Minkowski space 
the energy-momentum distribution of the matter fields is described by 
the symmetric energy-momentum tensor $T_{ab}$ satisfying the dominant 
energy condition. Then the quasi-local energy of the matter fields 
$E[D, K^a]$ with respect to some constant future pointing timelike 
unit vector field $K^a$ (i.e. a time translational Killing vector of 
the flat spacetime) is defined to be the integral $\int_D K_aT^{ab}t_b
{\rm d}\Sigma$ on the compact domain $D\subset\Sigma$ with boundary 
${\cal S}:=\partial D$ in some spacelike hypersurface $\Sigma$. Here 
$t^a$ is the future pointing unit normal to $\Sigma$ and ${\rm d} 
\Sigma$ is the natural volume element. As a consequence of the dominant 
energy condition this is not only positive definite, but also is a 
monotonically {\em increasing} set function: 
if $D_1\subset D_2$ then $E[D_1,K^a]\leq E[D_2,K^a]$. This implies, in 
particular, that in asymptotically flat configurations (when $\Sigma$ 
extends to spatial infinity and the total energy $E[\Sigma,K^a]$ is 
finite) $E[D,K^a]$ tends to the total energy {\em from below} as $D$ 
is blown up to exhaust $\Sigma$. 

Although in general relativity there is no well defined energy-momentum 
{\em density} of the gravitational `field', in asymptotically flat 
configurations its total (ADM) energy could be defined, and one of 
the greatest successes of classical general relativity in the last 
third of the 20th century is certainly the proof by Schoen and Yau 
\cite{ScYa} that the total gravitational energy is strictly positive 
definite. The logic of one of its simplest proofs, due to Witten 
\cite{Wi} (and simplified and corrected by Nester \cite{Ne}), is that 
we can rewrite the total energy as an integral of some expression 
(the so-called Sparling form \cite{Sp}, see also \cite{ReTo}) on a 
spacelike hypersurface, and by Witten's {\em gauge condition} the 
integrand could be ensured to be pointwise strictly positive definite. 
Thus the negative definite part of the Sparling form in the integrand 
is a pure gauge term. 

These results may yield the view that if the gravitational mass could 
be defined at the quasi-local level, then it would have to be not only 
positive definite, but also that in asymptotically flat spacetimes it 
would have to tend to the ADM mass as an {\em increasing} set function. 
(Note that while we can compare the quasi-local {\em masses}, i.e. 
scalars, we can compare the quasi-local {\em energies}, i.e. components 
of four-vectors on {\em different 2-surfaces} only in the presence of 
some extra structure, e.g. for spherically symmetric or stationary 
systems.) 
In fact, several specific quasi-local mass expressions exist which 
satisfy these requirements (viz. the Bartnik \cite{Ba} mass, the mass 
built from the Dougan--Mason \cite{DoMa} energy-momenta, and the 
Misner--Sharp energy \cite{MiSh} for spherically symmetric 
configurations), or at least the second of these in certain special 
spacetime configurations (the Hawking \cite{Ha} or the Geroch energies 
\cite{Ge} and the Penrose mass \cite{Pe}). However, there are other 
constructions (e.g. the Brown--York expressions \cite{BY}, the Epp 
\cite{E}, the Kijowski--Liu--Yau \cite{Ki,LY} and the Wang--Yau 
energies \cite{WY}) which tend to the ADM energy as {\em decreasing} 
set functions. (For a review and a more detailed discussion of the 
various quasi-local energy constructions and the extended literature, 
see e.g. \cite{szab04}.) 
These different monotonicity properties of the quasi-local mass/energy 
expressions generated some debate in the relativity community: 
whether a physically reasonable quasi-local energy expression should 
be monotonically {\em increasing} or {\em decreasing} near spatial 
infinity as it tends to the ADM energy. In fact, near spatial 
infinity the matter and the radiation can be neglected compared to 
the {\em negative definite} Newtonian gravitational binding energy. 
Therefore, increasing the domain of integration we have more and 
more {\em negative definite} contribution to the total energy, which 
therefore must be a decreasing set function (see also \cite{Niall}). 

Since near spatial infinity the dynamics of the fields and the 
gravitation dies off rapidly (typically as $1/r^2$), in the analysis 
of the asymptotic behaviour of any quasi-local energy/mass expression 
it is natural to consider the spacetime to be {\em static} in the 
first approximation. The advantage of the existence of a static 
Killing vector is that it provides a geometrically preferred notion 
of time and a preferred foliation of spacetime by a geometrically 
distinguished family of extrinsically flat spacelike hypersurfaces. 
In fact, the Newtonian limit of general relativity is defined in 
this way in \cite{HE}. Identifying the Newtonian potential $\phi$ 
with the logarithm of the length of the Killing field, we show that 
the exact field equations for this $\phi$ take the form of a Poisson 
equation in which the source term contains not only the familiar 
rest mass density of the matter fields (the Newtonian source term), 
but (among others) minus the {\em square} of the gradient of $\phi$ 
(post-Newtonian corrections) as well. Indeed, while the former is 
independent of $c^{-2}$, the latter is proportional to $c^{-2}$ and 
can naturally be interpreted as the energy-density (and/or the trace 
of the spatial stress tensor) of the gravitational field itself. Thus, 
to avoid confusion, we use the phrase `post-Newtonian' in the following 
sense: (1) the spacetime is static and asymptotically flat, in which (2) 
when quantities are expanded as a series of $c^{-2}$ then the zeroth 
order term are called Newtonian and the $c^{-2k}$ order ones the $k$th 
order post-Newtonian corrections. 

In the relativistically corrected Newtonian theory of gravity the 
volume integral of the source terms for the Newtonian potential 
gives a {\em well defined} (i.e. free of the ambiguities coming from 
the Galileo--E\"otv\"os experiment) expression for the energy of the 
source plus gravity system even at a quasi-local level. Motivated by 
this observation, we show that in static spacetimes in {\em exact 
general relativity} the extra structures above are enough to introduce 
a notion of `effective' quasi-local {\em energy} expression simply as 
the integral of the sum of all the source terms for $\phi$. 
This energy is shown to tend at spatial infinity to the ADM energy as 
a monotonically {\em decreasing} set function. We give an explicit 
form of this `effective' quasi-local energy in static, spherically 
symmetric spacetimes, and under certain energy conditions we prove its 
positivity and its rigidity, i.e. that its vanishing implies flatness. 

Since the idea behind the `effective' quasi-local energy expression 
is exactly analogous to the  quasi-local energy in the relativistically 
corrected Newtonian theory, we believe that the qualitative behaviour 
of the `effective' quasi-local energy expression reflects some 
universality: we expect that any physically acceptable quasi-local 
{\em energy} expression in a static, asymptotically flat spacetime 
should tend to the ADM energy as a {\em decreasing} set function at 
spatial infinity. 

In Section 2 we discuss the issue of energy both in Newtonian theory 
and in the relativistically corrected Newtonian theory. Then, in 
subsection 3.1, the Newtonian limit of Einstein theory is reviewed. 
We find that as a relativistic correction not only the energy density, 
but {\em the spatial stress of the gravitational field} also 
contributes to the effective source in the {\em exact} field equations 
for the Newtonian potential. We emphasize that this result is exact, 
i.e. no approximation is used. As far as we know this is a new 
post-Newtonian correction from GR, which has not been considered so 
far. Then the `effective' quasi-local energy is introduced and 
analyzed in subsections 3.2 and 3.3, and compared with the 
Misner--Sharp and Brown--York expressions in subsection 3.4. 
Section 4 is devoted to the discussion of the potential implications 
for more general quasi-local energy-momentum expressions. 

The sign conventions for the metric and the curvature of \cite{PR} are 
used. In particular, the signature of the spacetime metric is 
$(+,-,-,-)$, the curvature and Ricci tensors and the curvature scalar 
are defined by $-R^a{}_{bcd}X^bv^cw^d:=v^c\nabla_c(w^d\nabla_dX^a)-
w^c\nabla_c(v^d\nabla_dX^a)-[v,w]^c\nabla_cX^a$, $R_{bd}:=R^a{}_{bad}$ 
and $R:=R_{ab}g^{ab}$, respectively. Thus Einstein's equations take the 
form $G_{ab}:=R_{ab}-\frac{1}{2}Rg_{ab}=-\kappa T_{ab}-\lambda g_{ab}$, 
where $\lambda$ is the cosmological constant and $\kappa:=8\pi G/c^4$ 
with Newton's gravitational constant $G$ and the speed of light $c$; 
i.e. we use the traditional units. 


\section{Gravitational energy in Newton's theory}
\label{sec-2}

\subsection{Newton's theory}
\label{sub-2.1}

In a given inertial frame of reference the gravitational field is 
described by a scalar function $\phi$ of the flat 3-space $\mathbb{R}
^3$, for which the field equation is the Poisson equation 
\[
-D_aD^a\phi=4\pi G\rho.
\] 
Here $\rho:\mathbb{R}^3\rightarrow[0,\infty)$ is the 
{\em rest mass} density of the matter (source), and $D_a$ is the {\em 
flat} covariant derivative operator in the 3-space. Note that we use 
the negative definite flat metric $h_{ab}$ here, consistent with our 
conventions. If $D\subset\mathbb{R}^3$ is any open subset with compact 
closure and a smooth boundary ${\cal S}=\partial D$, then by the Gauss 
theorem and the field equation 

\begin{equation}
m_D:=\int_D\rho\,{\rm d}^3x=\frac{1}{4\pi G}\oint_{\cal S}v^e\bigl(D_e
\phi\bigr){\rm d}{\cal S}, \label{eq:2.1}
\end{equation}
where $v^e$ denotes the outward pointing unit normal of ${\cal S}$ and 
${\rm d}{\cal S}$ is the induced area element. Thus the {\em rest mass 
of the source} in Newtonian theory of gravity can be rewritten into a 
2-surface integral, like the charge in electrostatics. Following the 
analogy with electrostatics, we can introduce the {\em energy density} 
and the {\em spatial stress} of the gravitational field itself, 
respectively, by 

\begin{eqnarray}
U\!\!\!\!&:=\!\!\!\!&\frac{1}{8\pi G}h^{ab}\bigl(D_a\phi\bigr)\bigl(
 D_b\phi\bigr)=-\frac{1}{8\pi G}\vert D_a\phi\vert^2, 
 \label{eq:2.2.a} \\
\Sigma_{ab}\!\!\!\!&:=\!\!\!\!&\frac{1}{4\pi G}\Bigl(\bigl(D_a\phi
 \bigr)\bigl(D_b\phi\bigr)-\frac{1}{2}h_{ab}\bigl(D_c\phi\bigr)\bigl(
 D^c\phi\bigr)\Bigr). \label{eq:2.2.b}
\end{eqnarray}
In fact, the integral of $U$ on $\Sigma$ is just the work that we 
should do to form e.g. a spherical body by bringing particles 
together from infinity. Moreover, since gravitation is always 
attractive, in Newton's theory this is always a {\em binding} energy, 
and hence its sign is negative. The divergence of the stress tensor, 
together with the Poisson equation, yields the `force density': $D_a
\Sigma^{ab}=\frac{1}{4\pi G}(D_aD^a\phi)D^b\phi=-\rho D^b\phi$. Note 
also that the average `gravitational pressure' is just one-third of 
the gravitational energy density: $3P:=-h^{ab}\Sigma_{ab}=U$. 

However, by the Galileo--E\"otv\"os experiment there is an important 
difference between electrostatics and gravitation. Namely, by this 
experiment the inertial and gravitational masses of the particles 
are strictly proportional to each other. Hence, in particular, the 
gravitational and inertial masses of the test particles drop out 
from the equations of motion in the gravitational field, yielding 
an ambiguity even in the notion of the gravitational force $D_e\phi$: 
it is not possible, even in principle, to make a distinction between a 
{\em uniform} gravitational field and a {\em uniform} acceleration of 
the frame of reference. Therefore, at any {\em given} point of the 
3-space the gravitational force $D_e\phi$ can be transformed to any 
given value, e.g. to zero, by an appropriate change of the reference 
frame. Thus the ambiguity in the gravitational force is $D_e\phi
\mapsto D_e\phi+a_e$, where $a_e$ is an arbitrary {\em constant} 
covector field in 3-space. It is only the {\em second} derivative 
$D_aD_b\phi$, the tidal force, that has direct physical meaning. 
Consequently, the gravitational energy {\em density} (\ref{eq:2.2.a}) 
can also be transformed to any given non-positive value, e.g. to zero 
at any given point by an appropriate change of the frame of reference. 
Similarly, the spatial stress (\ref{eq:2.2.b}) is also vanishing at 
that given point. On the other hand, the gravitational energy density 
(as well as the spatial stress) can be transformed to zero on an 
extended, {\em open} subset of the 3-space {\em only if} the 
gravitational field is uniform there. If, however, we have some extra 
information about the structure of the gravitational field, e.g. that 
it is the gravitational field of a localized source, then the 
ambiguity can be removed from the gravitational force $D_a\phi$ by 
requiring its vanishing at infinity. 

To summarize, we see that even in Newtonian theory of gravity the 
gravitational energy and spatial stress {\em cannot be localized to a 
point}, just as a consequence of the Galileo--E\"otv\"os experiment, 
and the {\em rest mass of the source} can also be written as a {\em 
2-surface integral}. 


\subsection{Two relativistic corrections to the source}
\label{sub-2.2}

According to the special theory of relativity mass and energy are 
not independent concepts, and we should associate a mass distribution 
to any distribution of energy in 3-space. In particular, in addition 
to the mass distribution $u/c^2$ of the {\em internal energy} density 
$u$ of the matter field, we should associate a mass distribution 
$U/c^2$ to the energy density (\ref{eq:2.2.a}) of the gravitational 
field, too. However, according to the principle of equivalence, a 
consequence of the Galileo--E\"otv\"os experiment, {\em any} mass 
distribution is a source of gravity, independently of the nature of 
the mass. Thus, in particular, both the internal energy density of 
the matter and the gravitational energy density contribute to the 
source of gravity. Therefore, the source term on the right hand side 
of the Poisson equation should be corrected, and the field equation 
for the relativistically corrected Newtonian theory of gravity could 
naturally be expected to be 

\begin{equation}
-D_aD^a\phi=4\pi G\Bigl(\rho+\frac{1}{c^2}\bigl(u+U\bigr)\Bigr). 
\label{eq:2.3}
\end{equation}
Note that the gravitational energy density {\em reduces} the magnitude 
of the source, because that is a binding type energy. As a consequence 
of (\ref{eq:2.3}) we have that 

\begin{equation}
E_D:=\int_D\bigl(\rho c^2+u+U\bigr){\rm d}^3x=\frac{c^2}{4\pi G}\oint
_{\cal S}v^e\bigl(D_e\phi\bigr){\rm d}{\cal S}; \label{eq:2.4}
\end{equation}
i.e. now it is the {\em total energy of the source plus gravity 
system} in a given domain $D$ that can be rewritten into the form of a 
2-surface integral. Note that while the gravitational energy {\em 
density} is ambiguous, this {\em quasi-local} expression for the energy 
of the matter plus gravity system is {\em free of this ambiguity}, just 
because the flux integral on ${\cal S}$ of any constant covector field 
$a_e$ is zero. This in itself already justifies the introduction and 
use of the quasi-local concept of energy in the study of gravitating 
systems. Moreover, in the source-free region (i.e. where $\rho$ and 
$u$ are vanishing) $E_D$ is a {\em decreasing} set function, because 
then the integrand in the middle term is negative definite there. 
In particular, for a 2-sphere with radius $r$ surrounding a localized 
spherically symmetric homogeneous source, the quasi-local energy is 
$E_{D_r}=\frac{8\pi}{\kappa}{\tt m}(1+\frac{1}{2}\frac{\tt m}{r}+
\frac{{\tt m}^2}{r^2})+O(r^{-4})$, where, in terms of the rest mass $M$ 
and the radius $R$ of the source, the {\em total mass parameter} is 
${\tt m}=\frac{GM}{c^2}(1-\frac{3}{5}\frac{GM}{c^2R}+\frac{51}{140}
(\frac{GM}{c^2R})^2)+O(c^{-8})$.


\section{Gravitational energy in static spacetimes}
\label{sec-3}

\subsection{The Newtonian limit of Einstein's theory and two more 
relativistic corrections}
\label{sub-3.1}

In \cite{HE}, pp. 71--74, the Newtonian limit of Einstein's theory 
is defined through asymptotically flat, static configurations. Thus 
let the spacetime be static, and $K^a$ be the (e.g. future pointing) 
timelike Killing field being orthogonal to the spacelike level sets 
$t={\rm const}$ of a function $t:M\rightarrow\mathbb{R}$. These sets 
will be denoted by $\Sigma_t$, and we write $K_a=g\nabla_at$ for 
some function $g$ on $M$. If $f^2:=K_aK^a$, then $t^a:=f^{-1}K^a$ is 
the future pointing unit timelike normal to the hypersurfaces $\Sigma
_t$, and $h_{ab}:=g_{ab}-t_at_b$ is the induced (negative definite) 
metric on $\Sigma_t$. Let $D_a$ denote the corresponding intrinsic 
Levi-Civita covariant derivative operator on $\Sigma_t$. Then by a 
straightforward calculation it is shown in \cite{HE}, pp 72, that the 
length $f$ of the Killing field satisfies the `field equation' $h^{ab}
D_aD_bf=fR_{ab}t^at^b$. 

Next, let us define the energy density $\mu:=T_{ab}t^at^b$ of the 
matter fields seen by the static observers, decompose it into the sum 
of the rest mass and internal energy densities as $\mu=c^2\rho+u$, 
and write the trace of the energy-momentum tensor as $T_{ab}g^{ab}=
\mu-3p$. (Thus $-3p$ denotes the trace of the spatial stress tensor 
$\sigma_{ab}:=P^c_aP^d_bT_{cd}$ with respect to the {\em negative} 
definite $h_{ab}$. Here, $P^a_b$ denotes the obvious projection to 
$\Sigma_t$.) Introducing the scalar field $\phi:=c^2\ln f$, the field 
equation for $f$ above, together with Einstein's equations, gives 

\begin{equation}
-h^{ab}D_aD_b\phi=4\pi G\rho+\frac{4\pi G}{c^2}\Bigl(u+3p-\frac{c^4
\lambda}{4\pi G}\Bigr)+\frac{1}{c^2}h^{ab}\bigl(D_a\phi\bigr)\bigl(
D_b\phi\bigr). \label{eq:3.1}
\end{equation}
Comparing this equation with (\ref{eq:2.3}) we see that apparently 
we recovered (\ref{eq:2.3}) with the relativistic correction term 
$U/c^2$ dictated by the Galileo--E\"otv\"os experiment (even with 
the correct sign), together with an additional relativistic 
correction: the trace of the spatial stress (as well as the 
cosmological constant) also contributes to the effective source 
(as already noted in \cite{HE}). Nevertheless, the relative weight 
of the gravitational energy density term in the effective source 
is {\em twice} that of (\ref{eq:2.3}): the last term on the right 
hand side of (\ref{eq:3.1}) is $\frac{8\pi G}{c^2}U$ rather than 
the expected $\frac{4\pi G}{c^2}U$. However, as we saw in 
subsection \ref{sub-2.1}, we can associate with the Newtonian 
gravitational field not only energy density but also {\em spatial 
stress}, and the corresponding average pressure $P$ is one-third 
of the energy density. Thus the `extra' gravitational energy 
density in (\ref{eq:3.1}) can be written as $3P$, i.e. the last 
term of (\ref{eq:3.1}) has the form $\frac{4\pi G}{c^2}(U+3P)$. 
Hence there is a {\em fourth relativistic correction} to 
(\ref{eq:2.3}): the gravitational stress also contributes to the 
effective source of gravity. Note that this correction is obtained 
in the {\em exact} theory, independently of any approximation 
method. (In addition, the intrinsic geometry $(\Sigma_t,h_{ab})$, 
and hence the Laplacian $-h^{ab}D_aD_b$, is {\em not} flat. This 
can also be considered as an additional correction to (\ref{eq:2.3}), 
but it does not seem to be possible to formulate its deviation from 
the flat-space Laplacian of $\phi$ in a {\em gauge invariant} way.)

Another (and perhaps more direct) derivation of (\ref{eq:3.1}) could 
be based on the evolution parts $P^c_aP^d_b(G_{cd}+\kappa T_{cd}+
\lambda g_{cd})=0$ of Einstein's equations in the standard 3+1 
decomposition. If $\xi^a=Nt^a+N^a$ is an evolution vector field which 
is compatible with the foliation $\Sigma_t$, then, using the 
Hamiltonian constraint $t^at^b(G_{ab}+\lambda g_{ab}+\kappa T_{ab})
=0$, the evolution equations are equivalent to 

\begin{eqnarray}
\dot\chi_{ab}\!\!\!\!&=\!\!\!\!&N\Bigl(-{}^3R_{ab}+2\chi_{ac}\chi^c{}
 _b-\chi\chi_{ab}\Bigr)+L_{\bf N}\chi_{ab}-D_aD_bN+ \nonumber \\
\!\!\!\!&+\!\!\!\!&\lambda Nh_{ab}+\kappa N\Bigl(-\sigma_{ab}+\frac{1}
 {2}\sigma^c{}_ch_{ab}+\frac{1}{2}\mu h_{ab}\Bigr), \label{eq:3.2}
\end{eqnarray}
where ${}^3R_{ab}$ is the Ricci tensor of the intrinsic 3-metric $h
_{ab}$ and $L_{\bf N}\chi_{ab}$ is the Lie derivative of the extrinsic 
curvature $\chi_{ab}$ of $\Sigma_t$ along the shift vector field. 
Choosing the leaves $\Sigma_t$ of the foliation to be the 
hypersurfaces to which the Killing field $K^a$ is orthogonal the 
extrinsic curvature is vanishing, and choosing the evolution vector 
field to be the Killing field itself, the lapse will be the length of 
the Killing vector and the shift will be zero. Then (\ref{eq:3.2}) 
takes the form 

\begin{equation}
-D_aD_bf=f\Bigl({}^3G_{ab}+\kappa\sigma_{ab}+\frac{1}{2}\kappa(\mu+
3p)h_{ab}\Bigr), \label{eq:3.3}
\end{equation}
where ${}^3G_{ab}$ denotes the Einstein tensor of the spatial metric 
$h_{ab}$ and we used the Hamiltonian constraint 

\begin{equation}
\frac{1}{2}{}^3R=\kappa\mu+\lambda. \label{eq:3.4}
\end{equation}
(Since in static spacetimes the local momentum density $t^aT_{ab}P^b
_c$ of the matter fields is vanishing, the momentum constraint is 
satisfied identically.) Taking the trace of (\ref{eq:3.3}) and using 
$\phi$ instead of $f$ we recover (\ref{eq:3.1}). In what follows we 
need the full (\ref{eq:3.3}) rather than only its trace, and we 
consider (\ref{eq:3.3})-(\ref{eq:3.4}) to be the field equations 
rather than only (\ref{eq:3.1}). In fact, in the static case 
(\ref{eq:3.3})-(\ref{eq:3.4}) are equivalent to Einstein's equations, 
the field equations for the Newtonian potential $\phi$ and the spatial 
metric $h_{ab}$. 


\subsection{The `effective' quasi-local energy for static configurations 
and its spatial infinity limit}
\label{sub-3.2}

By the Galileo--E\"otv\"os experiment {\em any} kind of energy is a 
source of gravitation. Thus, motivated by expression (\ref{eq:2.4}) 
of the relativistically corrected Newtonian theory, in exact general 
relativity in static spacetimes it is natural to {\em define} the total, 
`effective' energy of the static matter+gravity system in a subset 
$D\subset\Sigma_t$, seen by the static observers $t^a$, as the integral 
of all the source terms on the right hand side of (\ref{eq:3.1}): 

\begin{equation}
E_D:=\int_D\Bigl(\mu+3p-\frac{c^4\lambda}{4\pi G}-\frac{1}{4\pi G}
\vert D_a\phi\vert^2\Bigr){\rm d}\Sigma=\frac{c^2}{4\pi G}\oint_{\cal 
S}v^a\bigl(D_a\phi\bigr){\rm d}{\cal S}. \label{eq:3.5}
\end{equation}
Note that $E_D$ contains not only the energy of the gravitational 
`field' and (all kinds of) energy of the matter source, but the trace 
of the spatial stress of the source and the gravitational `field' as 
well. Thus, apart from the cosmological term, the structure of the 
post-Newtonian part of the volume integral, $u+U+3(p+P)$, shows some 
resemblance to enthalpy rather than to the internal energy density. 
However, this combination seems to deviate from the standard form of 
enthalpy, too, which would have the structure $u+U+(p+P)$. (For a more 
detailed discussion of the analogy of gravitational energy with the 
thermodynamical ones, see \cite{Ki}.) 
If the source is compactly supported in some $D_0\subset D\subset\Sigma
_t$, and the cosmological constant is non-negative, then outside $D_0$ 
the total gravitational energy is strictly {\em decreasing} with 
increasing domain $D$ of integration. 

If the spacetime is asymptotically flat (in which case $\lambda=0$) 
and the hypersurfaces extend to the spatial infinity, then $E_{D_R}$, 
the quasi-local energy associated with the solid ball of radius $R$ 
(or equivalently to the sphere ${\cal S}_R=\partial D_R$), tends to 
the ADM energy in the $R\rightarrow\infty$ limit. To see this, it is 
enough to show that $E_{D_R}$ tends to Komar's expression because it 
is known that the Komar expression built from the static Killing field 
$K^a$ (normalized to one at infinity) tends to the ADM energy (see e.g. 
\cite{Beig}). Recall that Komar's integral on a closed spacelike 
2-surface ${\cal S}$ (with a timelike and spacelike unit normal, $t^a$ 
and $v^a$, respectively, and satisfying $t_av^a=0$) has the form 

\begin{equation}
{\tt I}_{\cal S}\bigl[K^e\bigr]:=\frac{1}{\kappa}\oint_{\cal S}\nabla
^{[a}K^{b]}\frac{1}{2}\varepsilon_{abcd}=\frac{2}{\kappa}\oint_{\cal S}
\bigl(v^at^b\nabla_{[a}K_{b]}\bigr){\rm d}{\cal S},  \label{eq:3.6}
\end{equation}
where ${\rm d}{\cal S}:=\frac{1}{2}t^ev^f\varepsilon_{efcd}$ is the 
induced area element on ${\cal S}$. Using the Killing equation and 
the form $K_a=\exp(\frac{\phi}{c^2})t_a$ of the Killing field, the 
integrand can be written as 

\begin{equation}
2v^at^b\nabla_{[a}K_{b]}=2v^a\bigl(\nabla_aK_b\bigr)
t^b-v^at^b\bigl(\nabla_aK_b+\nabla_bK_a\bigr)=2\exp\bigl(\frac{\phi}
{c^2}\bigr)v^aD_a\phi. \label{eq:3.7}
\end{equation}
Substituting this into (\ref{eq:3.6}) and taking into account that 
$\exp(\frac{\phi}{c^2})\rightarrow1$ at infinity,  we find that 
$\lim_{R\rightarrow\infty}E_{D_R}=\lim_{R\rightarrow\infty}{\tt I}_{{\cal S}
_R}[K^a]$.


\subsection{Static spherically symmetric configurations: Explicit 
form and positivity}
\label{sub-3.3}

Let the line element of the spatial 3-metric be written as $dh^2=-
e^{2\alpha}dr^2-R^2(d\theta^2+\sin^2\theta d\phi^2)$ for some 
functions $R$ and $\alpha$ of $r$ and regular at the origin $r=0$ 
with $R(0)=0$. Then the components of the outward pointing unit 
normal $v^a$ of the ${\cal S}_r:=\{r={\rm const}\}$ 2-surfaces are 
$v^e=e^{-\alpha}\delta^e_1$, and the corresponding extrinsic curvature 
is proportional to the induced 2-metric: $\nu_{ab}=\frac{R'}{R}e
^{-\alpha}(-R^2(\delta^2_a\delta^2_b+\sin^2\theta\delta^3_a\delta^3
_b))$. Here the prime denotes the derivative with respect to $r$. 
The curvature scalar of the spatial 3-metric is 

\begin{equation}
{}^3{\cal R}=\frac{2}{R^2}\Bigl(1+2RR' e^{-2\alpha}\alpha'-2RR''e
^{-2\alpha}-(R')^2e^{-2\alpha}\Bigr); \label{eq:3.8}
\end{equation}
while the curvature scalar of the intrinsic 2-metric on ${\cal S}_r$ 
is ${}^2{\cal R}=2/R^2$. (To avoid confusion, in this subsection the 
scalar curvatures are denoted by ${\cal R}$.) 

To give an explicit form of the `effective energy' in terms of the 
geometrical quantities defined on the 2-surface ${\cal S}_r$, we use 
not only the trace, but also the $v^av^b$ component of the evolution 
equation (\ref{eq:3.3}). The former is 

\begin{equation*}
0=D_aD^af+f\Bigl(\frac{1}{2}\kappa(\mu+3p)-\lambda\Bigr)=-e^{-\alpha}
\bigl(e^{-\alpha}f'\bigr)'-2\frac{R'}{R}e^{-2\alpha}f'+f\Bigl(\frac{1}
{2}\kappa(\mu+3p)-\lambda\Bigr),
\end{equation*}
while the latter is 

\begin{eqnarray*}
0\!\!\!\!&=\!\!\!\!&v^av^bD_aD_bf+f\Bigl({}^3G_{ab}v^av^b+\kappa\sigma
 _{ab}v^av^b-\frac{1}{2}\kappa(\mu+3p)\Bigr)= \nonumber \\
\!\!\!\!&=\!\!\!\!&e^{-\alpha}\bigl(e^{-\alpha}f'\bigr)'+f\Bigl(\frac{1}
 {R^2}-\bigl(\frac{R'}{R}\bigr)^2e^{-2\alpha}+\kappa T_{ab}v^av^b-
\frac{1}{2}\kappa(\mu+3p)\Bigr).
\end{eqnarray*}
Here, in the derivation of the second equation, we used the constraint 
(\ref{eq:3.4}) and the expression ${}^3G_{ab}v^av^b=\frac{1}{2}({}^2
{\cal R}+\nu_{ab}\nu^{ab}-\nu^2)$ for the $v^av^b$ component of the 
Einstein tensor of the 3-space in terms of the intrinsic and the 
extrinsic curvatures of ${\cal S}_r$. Comparing these two and assuming 
that $R'\not=0$, we obtain the expression of $f^{-1}f'$ in terms of $R$, 
$\alpha$ and $T_{ab}v^av^b$. However, this is essentially the integrand 
of the 2-surface integral in (\ref{eq:3.5}), $v^eD_e\phi=c^2e^{-\alpha}
f^{-1}f'$, yielding the explicit form of $E_r$ in spherically symmetric 
spacetimes: 

\begin{equation}
E_r=\frac{4\pi}{\kappa}\frac{R}{R'}e^\alpha\Bigl(R^2\bigl(\kappa T_{ab}
v^av^b-\lambda\bigr)+1-(R')^2e^{-2\alpha}\Bigr). \label{eq:3.9}
\end{equation}
This formula can be simplified slightly if we choose $r$ to be the 
areal coordinate, i.e. $r=R$, and hence $R'=1$. (\ref{eq:3.9}) gives 
$E_r$ {\em algebraically} in terms of the functions in the 3-metric 
and the radial pressure $T_{ab}v^av^b$ {\em on the surface} ${\cal S}
_r$ of radius $r$. 

In the rest of this subsection we prove that this energy expression 
is non-negative if the matter fields satisfy the `energy conditions' 
$(\kappa T_{ab}+\lambda g_{ab})t^at^b\geq0$ and $(\kappa T_{ab}+\lambda 
g_{ab})v^av^b\geq0$. Moreover, we show that, under the slightly 
stronger conditions $(\kappa T_{ab}+\lambda g_{ab})t^at^b\geq(\kappa T
_{ab}+\lambda g_{ab})v^av^b\geq0$, the vanishing of $E_r$ implies 
$\kappa T_{ab}=-\lambda g_{ab}$ and the flatness of the Cauchy 
development of the ball with radius $r$. Clearly, $(\kappa T_{ab}+
\lambda g_{ab})v^av^b\geq0$ ensures the non-negativity of the first 
term between the brackets in (\ref{eq:3.9}). To show that the second 
term dominates the third we use the constraint equation 
(\ref{eq:3.4}). Substituting (\ref{eq:3.8}) into (\ref{eq:3.4}) and 
multiplying by $R^2R'$, we obtain $(R(R')^2e^{-2\alpha})'=R'-R^2R'
(\kappa\mu+\lambda)$. Integrating this from zero to $r$ and using 
$(\kappa T_{ab}+\lambda g_{ab})t^at^b\geq0$, we obtain 

\begin{equation}
R\bigl(R'\bigr)^2e^{-2\alpha}=R-\int^r_0\bigl(\kappa\mu+\lambda\bigr)
R^2R'dr\leq R; \label{eq:3.10}
\end{equation}
i.e. that $(R')^2\leq e^{2\alpha}$, and hence that $E_r\geq0$. (Here we 
used our previous assumption that $R$ is strictly monotonically 
increasing, i.e. no minimal or maximal 2-surface is present.) 

Conversely, by the pointwise non-negativity of $\kappa T_{ab}v^av^b-
\lambda$ and of $e^\alpha-(R')^2e^{-\alpha}$ from $E_r=0$ it follows 
their vanishing {\em on the 2-surface} ${\cal S}_r$, i.e. in 
particular that $R'=e^\alpha$. Substituting this to (\ref{eq:3.10}) 
we obtain that $\kappa\mu+\lambda=0$ {\em on the whole 3-ball of 
radius $r$}, and hence that $R'=e^\alpha$ also on the whole 3-ball 
(and not only on its boundary). Substituting this into the line 
element it becomes flat, i.e. the initial data set on the ball of 
radius $r$ is the trivial one, and hence its Cauchy development in 
the spacetime is also flat. Finally, by the stronger energy 
condition, $\kappa T_{ab}t^at^b=-\lambda$ implies $\kappa T_{ab}v^a
v^b=\lambda$, i.e. $\kappa T_{ab}=-\lambda g_{ab}$. 


\subsection{Comparison with other round sphere expressions}
\label{sub-3.4}

Since the Misner--Sharp energy appears as a mass expression in the 
study of equilibrium states of cold, spherically symmetric stars (see 
e.g. Appendix 1 of \cite{HE}), this became the more or less generally 
accepted definition of quasi-local energy on round spheres (i.e. on 
spherically symmetric 2-surfaces in spherically symmetric spacetimes). 
(Note that by spherical symmetry the spatial part of the 
energy-momentum is expected to be zero, and hence the mass is equal 
to energy.) In the line element of the previous subsection this takes 
the form $E_{MS}(r)=\frac{4\pi}{\kappa}r(1-e^{-2\alpha})$, where $r$ 
is the {\em areal} coordinate; while the Brown--York energy is $E
_{BY}(r)=\frac{8\pi}{\kappa}r(1-e^{-\alpha})$. 
(N.B.: On round spheres the Bartnik, the Dougan--Mason and the 
Penrose masses and the Hawking, the Geroch and the Kijowski energies 
reduce to $E_{MS}(r)$. On the other hand, the Brown--York energies 
(with all three choices for the reference configurations), the Epp, 
the Kijowski--Liu--Yau (which is Kijowski's free energy) and the 
Wang--Yau expressions reduce to $E_{BY}(r)$. For the details see e.g. 
\cite{szab04}.) Since the Hawking energy is a gauge invariant measure 
of the bending of the light rays orthogonal to the 2-surface (see 
subsection 6.1.1 of \cite{szab04}), the Misner--Sharp energy can also 
be interpreted in this way. 

On the other hand, the `effective' quasi-local energy was introduced 
as the integral of the effective source for the Newtonian potential 
{\em seen by the static observers}, i.e. there is a {\em different 
concept of energy} behind this: it is a measure of the effective 
source of gravity, including also the gravitational self-interaction. 
This yields that, in addition to the extra pressure and cosmological 
terms in (\ref{eq:3.9}), it has an extra overall weight function 
$e^\alpha$ with respect to $E_{MS}(r)$. Since by the positivity proof 
above it is not less than 1, the energy $E_r$ is never less than the 
Misner-Sharp energy. 

In the Schwarzschild solution the Misner--Sharp energy is the 
constant $\frac{8\pi}{\kappa}{\tt m}$ for any $r>2{\tt m}$, the 
`effective' quasi-local energy is $E_r=\frac{8\pi}{\kappa}\frac{\tt m}
{\sqrt{1-\frac{2{\tt m}}{r}}}$, while the Brown--York energy is $E
_{BY}(r)=\frac{8\pi}{\kappa}r(1-{\sqrt{1-\frac{2{\tt m}}{r}}})$. Thus, 
the Schwarzschild mass parameter ${\tt m}$ is only the `bare' mass, 
while in $E_r$ this `bare' mass is `dressed' by the inverse local 
redshift factor and it tends to the ADM energy from above. Similarly, 
$E_{BY}(r)$ is also monotonically decreasing. To see their more 
detailed asymptotic structure let us expand them as a power series of 
$1/r$ near infinity. We find 

\begin{eqnarray}
E_{BY}\bigl(r\bigr)\!\!\!\!&=\!\!\!\!&\frac{8\pi}{\kappa}{\tt m}\Bigl(
 1+\frac{1}{2}\frac{\tt m}{r}\Bigr)+O\bigl(r^{-2}\bigr), 
 \label{eq:3.11.a}\\
E_r\!\!\!\!&=\!\!\!\!&\frac{8\pi}{\kappa}{\tt m}\Bigl(1+\frac{\tt m}
 {r}\Bigr)+O\bigl(r^{-2}\bigr); \label{eq:3.11.b}
\end{eqnarray}
which, by ${\tt m}\sim Gc^{-2}$ (see the last sentence of subsection 
\ref{sub-2.2}), are the post-Newtonian expansions at the same time. 
Thus the Brown--York energy reproduces the Newtonian energy expression 
even in the $c^{-2}$ order. On the other hand, by general relativity 
it is {\em twice} the Newtonian gravitational energy (more precisely 
the sum of the Newtonian gravitational energy and the average pressure) 
that appears as the source of gravity in the first post-Newtonian order 
(see (\ref{eq:3.1}) and the subsequent discussion). The factor 2 in 
front of the post-Newtonian energy term in (\ref{eq:3.11.b}) is simply 
a manifestation of the contribution of the gravitational spatial 
stress. Since, however, $E_r$ is a measure of the source {\em seen by 
the static observers}, it diverges at the horizon as it could be 
expected. Thus our `effective' quasi-local energy may provide a 
physically reasonable notion of energy in the region where the 
Killing vector is timelike, i.e. outside the event horizon.


\section{Discussion}
\label{sec-4}

In the literature there exist lists of {\em a priori} expectations 
on how a physically reasonable quasi-local mass or energy-momentum 
expression should behave \cite{szab04,CY}, e.g. at spatial infinity, 
in the presence of spherical symmetry, or on the event horizon of 
black holes. 

One such additional natural expectation could be the compatibility 
with the results in the relativistically corrected Newtonian theory, 
where the quasi-locally defined energy of the matter plus gravity 
system tends to the total energy, measured at infinity, as a {\em 
strictly decreasing} set function. Thus, to be able to make this 
comparison, we need to define the Newtonian limit of general 
relativity. One of the several possible definitions is based on 
static spacetimes that are asymptotically flat at spatial infinity 
\cite{HE}. Then we can expand every quantity and formula as a series 
of $c^{-2}$, and while the zeroth order terms give the Newtonian 
approximation, the coefficients of $c^{-2k}$ are the $k$th order 
post-Newtonian corrections. In fact, since near spatial infinity in 
an asymptotically flat spacetime the matter fields and the dynamics 
of both the matter and the geometry die off rapidly, the quasi-local 
energy-momentum/mass expressions could be expected to behave like in 
{\em static} asymptotically flat spacetimes. Though this notion of 
post-Newtonian approximation is more restrictive than the usual one 
(see \cite{wald}), the advantage of this is that the presence of the 
{\em static} Killing field provides extra geometric structures that 
make the subsequent analysis technically much easier and unique. In 
particular, they make it possible to define {\em energy} at the 
quasi-local level and to be able to compare energies (and not only 
masses) associated with {\em different} 2-surfaces unambiguously. 

Since in static spacetimes the gravitational contribution to the 
total energy is {\em negative definite}, the quasi-local expressions 
should tend to the ADM expression as {\em strictly decreasing} 
functions. Therefore, in particular, if $E(r)$ is any such 
gravitational energy expression evaluated on a large sphere of radius 
$r$ near spatial infinity and $E_r$ denotes the `effective' energy, 
also at $r$, then the asymptotic form of the former could be expected 
e.g. to be $E(r)=E_r+O(r^{-2})$ or $E(r)=E_{BY}(r)+O(r^{-2})$, 
depending on the concept of energy that is behind the actual notion 
$E(r)$ of quasi-local energy. 

However, this new requirement is in conflict with two of the previous 
ones in \cite{szab04,CY}. First, this contradicts to the expectation 
that for round spheres the quasi-local energy should reduce to the 
Misner-Sharp energy since the latter is {\em increasing} (or constant). 
Thus in static spherically symmetric configurations $E_r$ could be an 
alternative to the Misner--Sharp expression. 

Second, if the quasi-local mass should really tend to the ADM mass as 
a {\em strictly decreasing} set function near spatial infinity, then 
the Schwarzschild example shows that the quasi-local mass at the event 
horizon {\em cannot} be expected to be the irreducible mass. In fact, 
since both the ADM and irreducible masses are $\frac{8\pi}{\kappa}
{\tt m}$ and the quasi-local mass must be strictly decreasing, there 
would have to be a closed 2-surface between the horizon and the 
spatial infinity on which the quasi-local mass would take its maximal 
value. However, it does not seem why such a (geometrically, and hence, 
physically) distinguished 2-surface should exist. 


\bigskip

This work was partially supported by the Hungarian Scientific Research 
Fund (OTKA) grant K67790 and by the Royal Society of New Zealand with 
Marsden grant UOO-09-022.


\end{document}